\documentclass{elsarticle}
\usepackage{mathtools}

\DeclarePairedDelimiter\floor{\lfloor}{\rfloor}

\usepackage{amssymb}

\usepackage[figuresright]{rotating}

\begin{document}

\begin{frontmatter}




\title{Low-lying energy bands in a finite periodic multiple-well potential}


\author{Dae-Yup Song}
\ead{dsong@sunchon.ac.kr}
\address{Department of Physics Education, Sunchon National University, Jeonnam 57922, Korea}

\begin{abstract}We  analyze the low-lying states for a one-dimensional potential consisting of $N$ identical wells, assuming that the wells are parabolic around the minima. Matching the exact wave functions around the minima and the WKB wave functions in the barriers, we find a quantization condition which is then solved to give a formula for the energy eigenvalues explicitly written in terms of the potential.  In addition, constructing $N$ localized approximate  eigenstates  each of which matches on to that of the harmonic oscillator in one of the parabolic wells,  and diagonalizing the Hamiltonian in the subspace spanned by the localized states on the assumption that  the localized states form an orthogonal basis,  we also find the same formula for the  energy eigenvalues which the method of matching the wave functions gives. In the large-$N$ limit, the formula reproduces,  at the leading order,  the expression for the  widths of the narrow energy bands of the Mathieu equation present in the mathematical literature. 
As there are differences between  the $N$-well system in the large-$N$ limit and the fully periodic system, we include a two-dimensional model in which the quadratic minima are located on the vertices of a regular $N$-sided polygon with rotational symmetry of order $N$. We argue that  the lowest band of the two-dimensional model closely resembles the  tight-binding energy bands of the  fully periodic one-dimensional system in that  most of the eigenvalues are degenerate in the large-$N$ limit with the eigenfunctions satisfying the Bloch condition under the discrete rotations. 
\end{abstract}


\end{frontmatter}



\section{ Introduction}
\label{Introduction}


Quantum tunneling is of continued interest since the advent of quantum mechanics. In addition to the well-known phenomena of microscopic quantum tunneling \cite{Merzbacher},
recently, macroscopic quantum tunneling has been realized in superconducting quantum interference device with Josephson junctions  (see, e.g., Refs.~\cite{Johnson,Koch}), and,
for the system of the transmon Hamiltonian of a cosine potential with the amplitude of the potential being large, the analytic expression for the energy splittings between the nearly degenerate even and odd eigenstates is given through a semiclassical approach \cite{CSDG}, which amounts to reproducing the rigorous mathematical expression for the widths of the low-lying energy bands of the  Mathieu equation at the leading order (see, e.g.,  Refs.~\cite{nist,CUMS}). For the derivation of the energy splitting in the periodic potential, the tight-binding approach (see, e.g.,  Ref.~\cite{Harrison}) is used, with the  localized semiclassical wave function which  matches the normalized wave function of a harmonic oscillator in the forbidden region   of the well of localization \cite{CSDG}.

It has long been recognized that the calculation of the energy splittings in a  symmetric double-well potential is closely related to that of the widths of the low-lying energy bands of a periodic potential \cite{CDL,Coleman}. For a symmetric double-well potential, if an approximate energy eigenstate localized in one well is assumed in the limit of low probability for barrier penetration, the state localized in the other well can also be obtained relying on the reflection symmetry of the system, so that the Hamiltonian is represented by a  $2\times 2 $ symmetric matrix in the subspace spanned by the two localized states on the assumption that  the two states form an orthonormal basis (see, e.g.,  Refs.~\cite{DekkerPRA,Song}). On the other hand, for the system of a periodic potential, if an approximate energy eigenstate localized in a well is assumed similarly, an infinite number of the localized states can  be obtained relying on the translational symmetry of the system, so that, in the subspace spanned by the localized states, the Hamiltonian is represented by an infinite-dimensional  matrix. As detailed in Ref.~\cite{CDL}, if the nearest-neighbor 
contributions dominate among the tunneling effect, the Hamiltonian matrix can be approximated as an infinite symmetric tridiagonal (Toeplitz) matrix on the assumption that  the localized states form an orthonormal basis. The eigenvalues of the infinite matrix can then be found by invoking the Bloch theorem \cite{CDL}, which formally agree with those obtained through the rigorous application of the tight-binding approximation \cite{Harrison} at the leading order. 

Recently, Sacchetti in his study of nonlinear Schr\"{o}dinger equation considers a $N$-well potential which is defined to be the same single-well potential repeated $N$ times \cite{Sacchetti}: In one dimension, the $N$-well potential is such that it coincides, on a finite domain of length $Na$, with the fully periodic potential consisting of the infinite repetition of the single-well potential,  where the period $a$ is the distance between the minima of the adjacent wells. In the linear system which we are interested in, it is also assumed that the  non-degenerate ground state of the single-well potential  may be considered as an approximate eigenstate of the $N$-well system \cite{Sacchetti}.  In the subspace spanned by the localized states which are obtained from the ground state relying on the finite periodicity, the  Hamiltonian is then represented by a $N\times N$ symmetric tridiagonal Toeplitz matrix, {\em again, on the assumption that} the $N$ states  form an orthonormal basis. The energy eigenvalues of the $N$-well system found in this "$N$-mode approximation" \cite{Sacchetti} turn out to be exactly those that the eigenvalue formula of the fully periodic system gives  at some discrete Bloch wavenumbers \cite{CDL} (see Section \ref{sec:TBA}).  If "Bloch phase" is defined to be Bloch wavenumber multiplied by $a$, we note that the discrete Bloch phases of Ref.~\cite{Sacchetti} already appear in the finite-periodic system consisting of $N$ square wells when the single square well is deep and wide \cite{SSWM}. We also note that some other discrete Bloch phases have long been found for the unit transmission in the scattering by the finite periodic potentials \cite{VC,SWM,GS,PP}.

Though the $N$-mode(level) approximation may be justified as a variational method \cite{CDL}, in order to find  the expressions of the off-diagonal (hopping) matrix element and thus of the eigenvalues  explicitly written in terms of the potential in this approximation, however, we may need some scheme with the  judiciously normalized  approximate wave functions \cite{CUMS,DekkerPRA,Song}.  For a double-well potential, instead of normalizing the localized states and applying the two-level approach with the Lifshitz-Herring approximation, Dekker first shows that, if we assume the two wells are parabolic,  the consistency condition which comes from that a WKB function should match onto the exact solutions of the parabolic cylinder functions on both sides of the potential barrier determines the energy splitting between the pair of the lowest eigenvalues \cite{Dekker}. While this Dekker's method exactly reproduces the known energy splitting formula which the instanton method \cite{Coleman} or the two-level approximation gives, it has been argued that the two-level approach could supplement the validity of Dekker's method in the vanishing limit of the energy splitting \cite{Song15}.

In this article, we will consider the  finite periodic $N$-well potential by assuming that the wells are parabolic. Analyzing the relations for the coefficients introduced through the application of  Dekker's method, we arrive at a quantization condition  for the  eigenvalues  in the limit of low probability for barrier penetration. The quantization condition is then solved to find a formula for the bands of  energy eigenvalues, where each band is associated with one of the low-lying states of the  harmonic oscillator of the single well. For the lowest band, the formula reproduces  the eigenvalues of Ref.~\cite{Sacchetti}, but, here, they are written explicitly in terms of the potential. 
As the validity of Dekker's method may not be provided for a case in which  the shift from one of the harmonic oscillator eigenvalues vanishes \cite{Song15}, we also apply  the $N$-level approach  with some approximations,
to find  the same formula for the energy eigenvalues that Dekker's method gives. 
As an application, it is  shown that the rigorous expression for the widths of the low-lying energy bands of the  Mathieu equation can be obtained  in the large-$N$ limit.  The eigenvalues in the system of the  one-dimensional finite periodic $N$-well potential, however, are not degenerate, which is a different feature from the tight-binding bands, 
and thus, in relation with the  $N=4$ case  of Ref.~\cite{Sacchetti} (see, also, Ref.~\cite{Kevrekidis}), we also consider a two-dimensional potential  in which the quadratic minima are located on the vertices of a regular $N$-sided polygon with rotational symmetry of order $N$. Through the $N$-level approximation,  it is shown that,  the lowest band of the two-dimensional model closely resembles the  tight-binding energy bands of the  fully periodic system \cite{Harrison,CDL},
in that  most of the eigenvalues are degenerate in the large-$N$ limit with the eigenfunctions satisfying the  Bloch condition under the discrete rotations. 

This paper is organized as follows: In Section \ref{sec:wave}, the exact wave functions around the minima of the wells and WKB wave functions in the barriers are introduced, and their asymptotic expansions are given. In Section \ref{sec:qcondition}, matching the wave functions in the overlapping regions, the linear relations between coefficients introduced for the wave functions are given, and a quantization condition is found as a consistency relation: The condition is then solved to give a formula for the energy eigenvalues.   In Section \ref{sec:N-level}, constructing the localized states, the $N$-level approximation is used to re-obtain the formula for the eigenvalues. In Section \ref{sec:comparison}, within the ``strong bonding approximation'' of Ref.~\cite{CDL}, the localized states are used in finding the energy eigenvalues of low-lying energy bands of the fully periodic system in terms of the potential and the  widths of the narrow energy bands of the Mathieu equation are calculated.  In Section \ref{sec:concluding}, we give some concluding remarks. Finally in  \ref{sec:appendix}, a two-dimensional model is introduced and explored in relation with the Bloch theorem. 

\section{ Exact solutions  and  WKB wave functions }
\label{sec:wave}
\begin{figure}
\centering
\includegraphics[width=11cm]{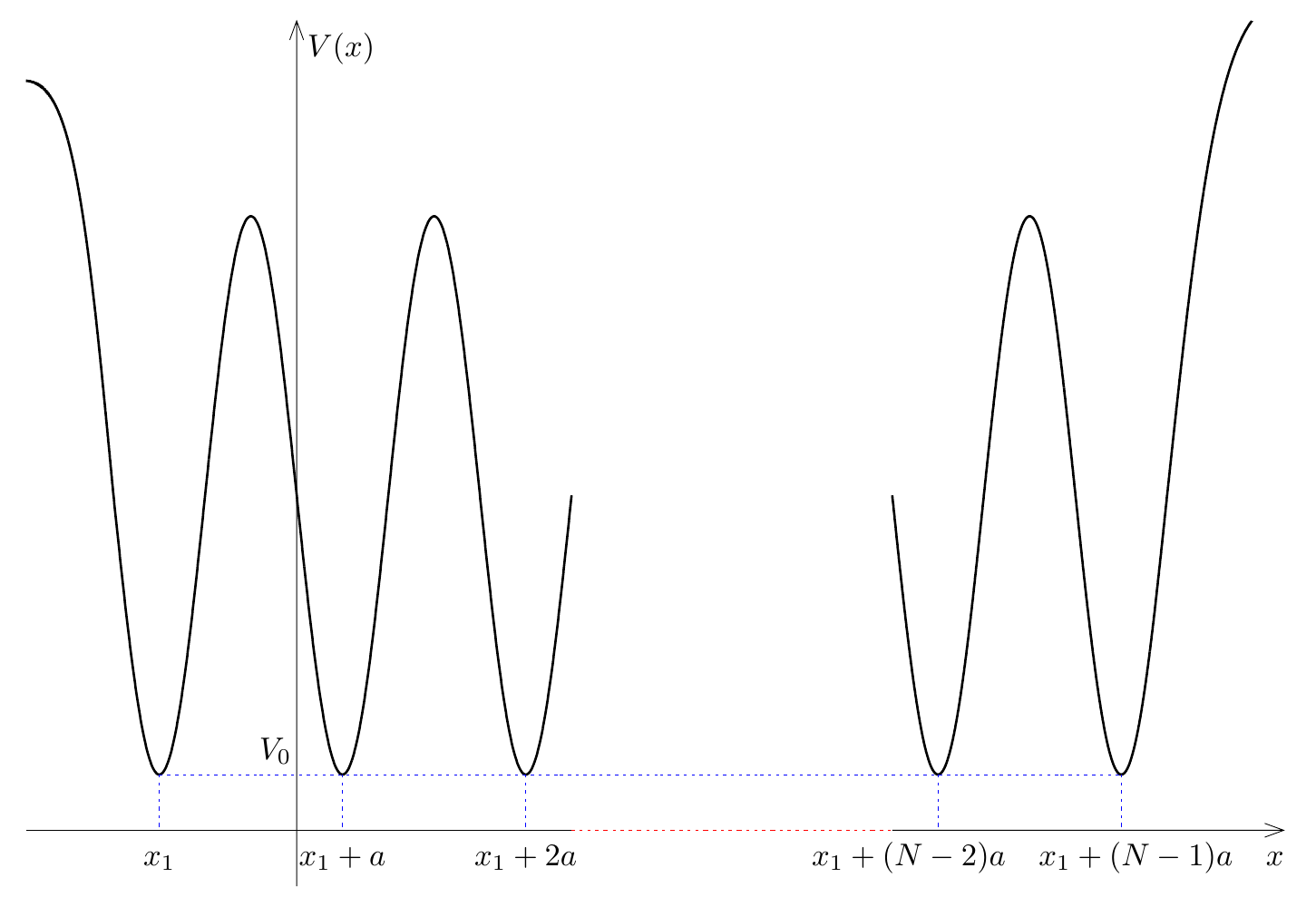}
\caption{  A finite periodic $N$-well potential $V(x)$. We assume that $V(x)$ is quadratic around its minima at $x=x_1+ja$ ($j=0,1, \ldots, N-1$) with  angular frequency $\omega$. }
\label{figure1}
\end{figure}

We assume the finite periodic $N$-well potential $V(x)$ satisfying $V(x+a)=V(x)$ with a positive constant $a$ for $x_1\le x \le  x_1+ (N-  2)a$, while smooth $V(x)$ is monotonically decreasing (increasing) for $x<x_1$ (for $x>x_1+(N-1)a$) (see Fig.~\ref{figure1}) so that low-lying energy spectrum is discrete with square-integrable eigenfunctions. We also assume that $V(x)$ has  quadratic minima at  $x=x_1+ja$  ($j=0,1,  \ldots, N-1$), and thus, in the quadratic region near $x=x_1+ja$, the potential is written as
\begin{equation}
V(x)= V_0 + {m\omega^2 \over  2} (x-x_1-ja)^2 ,
\end{equation}
with the particle's mass $m$ and angular frequency $\omega$.
For the eigenfunction $\psi(x)$ corresponding to the eigenvalues 
\begin{equation}
E(\nu)=V_0+(\nu +\frac{1}{2})\hbar\omega,
\label{Enu2}
\end{equation}
the Schr\"{o}dinger equation
\begin{equation}
H  \psi(x)  =-\frac{\hbar^2}{2m} \frac{d^2}{dx^2}\psi(x) +V(x) \psi(x)= E(\nu)\psi(x)
\end{equation}
is then written in the quadratic region  near $x=x_1+ja$ as
\begin{equation}
-\frac{\hbar^2}{2m} \frac{d^2}{dx^2}\psi(x) +\frac{m\omega^2}{2}(x-x_1-ja)^2\psi(x)= \hbar\omega (\nu+ \frac{1}{2})\psi(x).
\label{seqj}
\end{equation}
By introducing
\begin{equation}
 z_j=\frac{\sqrt{2}(x-x_1-ja)}{l};~~j=0,1,\ldots,N-1
\end{equation}
with
\begin{equation}
l=\sqrt{{\hbar}\over{m\omega}},
\end{equation}
we rewrite Eq.~($\ref{seqj}$)  as
\begin{equation}
\label{pcfrl}
\frac{d^2 \psi}{d z_j^2}+ \left(\nu+\frac{1}{2} -\frac{z_j^2}{4}\right)\psi =0.
\end{equation}
The solutions of Eq. (\ref{pcfrl}) are parabolic cylinder  functions, and we write the wave function $\psi(x)$, near $x=x_1+ja$ with $j=1,2,\ldots,N-2$,  as
\begin{eqnarray}
\psi_j(x)&=&C_j^+ D_{\nu} \left(   \sqrt{2}(x-x_1-ja)/{l}  \right) +C_j^- D_{\nu} \left( -  \sqrt{2}(x-x_1-ja)/{l}  \right)  \cr
&=&C_j^+ D_{\nu}( z_j)  +  C_j^- D_{\nu}(- z_j),
\label{pcfj}
\end{eqnarray}
with constant $C_j^+$ and $ C_j^-$.
On the other hand, as was first done by Dekker \cite{Dekker}, near  $x=x_1$  and near $x=x_1 +(N-1)a$, with  constant $C_0$ and $ C_{N-1}$ we write the wave function $\psi(x)$ as
\begin{equation}
\psi_0(x)= C_0 D_{\nu}(- z_0) ~~{\rm and}~~\psi_{N-1} (x)=C_{N-1} D_{\nu}( z_{N-1}),
\label{pcf0N-1 }
\end{equation}
respectively, bearing in mind that we wish to construct a normalizable wave function so that $\int_{-\infty}^{x_1} |\psi_0(x)|^2 dx$ ($\int_{x_1+(N-1)a}^\infty |\psi_{N-1}(x)|^2 dx$)  is finite  if we suppose the expression of $\psi_0(x)$ ($\psi_{N-1}(x)$) is valid for $x< x_1$ ($x>x_1+(N-1)a$).

\subsection{Asymptotic expansions of the exact solutions}
For real and positive $z$ satisfying $z\gg|\nu|$, we have the asymptotic expansion \cite{WW}:
\begin{equation}
D_\nu (z) \sim z^\nu \exp\left(-\frac{z^2}{4}\right)\left[1-\frac{\nu(\nu-1)}{2z^2}+\cdots\right],
\label{AsymDp}
\end{equation}
while, for real and negative $z$ satisfying $|z|\gg|\nu|$, the expansion is \cite{MG,AS,Song15}
\begin{eqnarray}
D_\nu (z)~ & \sim&~ \cos(\nu\pi)|z|^\nu\exp\left(-\frac{z^2}{4}\right)\left[1-\frac{\nu(\nu-1)}{2z^2}+\cdots\right]
\nonumber\\
&&~-\frac{\sin(\nu\pi)\nu !}{|z|^{\nu+1}} \sqrt{ \frac{2}{\pi}}\exp\left(\frac{z^2}{4}\right)
\left[1+\frac{(\nu+1)(\nu+2)}{2z^2}+\cdots\right].
\label{AsymDn}
\end{eqnarray}
In the quadratic region near $x=x_1$, if $z_0\gg |\nu|$, we thus have
\begin{equation}
\psi_0(x) \simeq  C_0 \left[ \cos(\nu\pi)z_0^\nu \exp\left(-\frac{z_0^2}{4}\right) 
   -\frac{\sin(\nu\pi)\nu !}{z_0^{\nu+1}} \sqrt{ \frac{2}{\pi}}\exp\left(\frac{z_0^2}{4}\right)\right].
\label{z0ggnu}
\end{equation}

For  $j=1,2,\ldots,N-2$, in the quadratic region near $x=x_1+ja$, using Eqs.~(\ref{pcfj}), (\ref{AsymDp}) and (\ref{AsymDn}) we find, for $z_j\gg|\nu|$,
\begin{eqnarray}
\psi_j(x) &\simeq& \left[ C_j^+ + C_j^- \cos(\nu\pi)\right] z_j^\nu \exp\left(-\frac{z_j^2}{4}\right)  \cr
&&-  C_j^-\frac{\sin(\nu\pi)\nu !}{z_j^{\nu+1}} \sqrt{ \frac{2}{\pi}}\exp\left(\frac{z_j^2}{4}\right), 
\label{zjggnu}
\end{eqnarray}
and, for $-z_j\gg|\nu|$, 
\begin{eqnarray}
\psi_j(x)
& \simeq& \left[ C_j^- + C_j^+ \cos(\nu\pi)\right] |z_j|^\nu \exp\left(-\frac{z_j^2}{4}\right)\cr
&&-  C_j^+\frac{\sin(\nu\pi)\nu !}{|z_j|^{\nu+1}} \sqrt{ \frac{2}{\pi}}\exp\left(\frac{z_j^2}{4}\right).
\label{mzjggnu}
\end{eqnarray}
In the quadratic region near $x=x_1+(N-1)a$, if $-z_{N-1}\gg |\nu|$, we  have
\begin{eqnarray}
\psi_{N-1}(x)& \simeq& C_{N-1} \cos(\nu\pi)|z_{N-1}|^\nu \exp\left(-\frac{z_{N-1}^2}{4}\right) \cr
&&  -C_{N-1}\frac{\sin(\nu\pi)\nu !}{|z_{N-1}|^{\nu+1}} \sqrt{ \frac{2}{\pi}}\exp\left(\frac{z_{N-1}^2}{4}\right).
\label{zN-1ggnu}
\end{eqnarray}

\subsection{ WKB wave functions  and the asymptotic expansions }
\begin{figure}
\centering
\includegraphics[width=11cm]{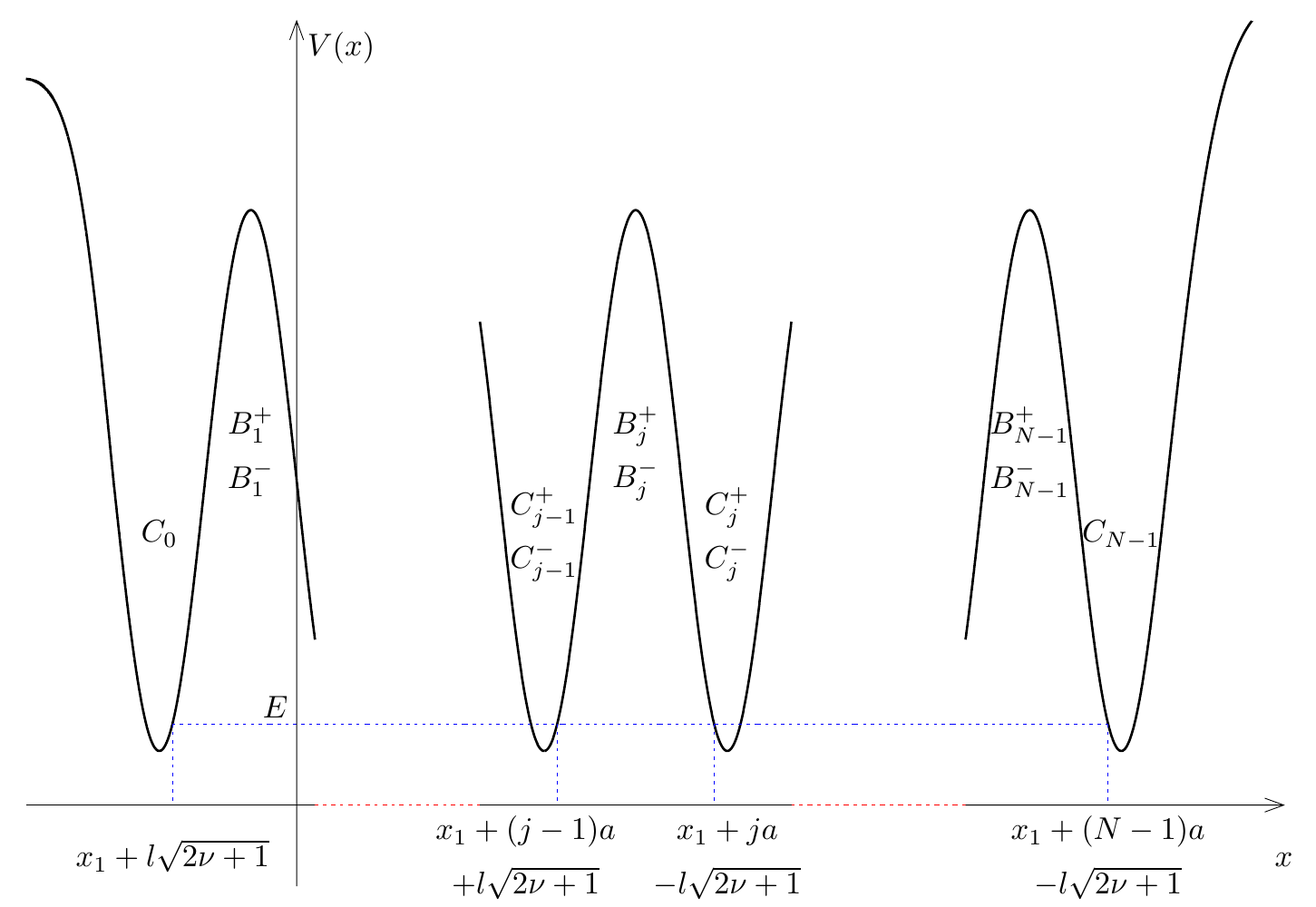}
\caption{  The turning points near  $x=x_1+(j-\frac{1}{2})a$, and at  $x=x_1+l\sqrt{2\nu+1},~ x_1+(N-1)a-l\sqrt{2\nu+1}$. The coefficients for the wave functions are  depicted in the corresponding regions of $x$. }
\label{figure2}
\end{figure}

We assume $a\gg l$ and  $x= x_1+ (j-\frac{1}{2})a $ is in the classically forbidden region, with the classical turning points at $x=x_1+ (j-1)a +l\sqrt{2\nu+1}$, $x_1+ja -l\sqrt{2\nu+1}$ satisfying (see Fig.~\ref{figure2})
\begin{equation}
V(x_1+( j-1)a +l\sqrt{2\nu+1})=V( x_1+ja -l\sqrt{2\nu+1})=E.
\end{equation}
In the  forbidden region near $x=x_1+ (j-\frac{1}{2})a$,
the WKB approximation to an eigenfunction is written as
\begin{eqnarray}
\psi_j^{WKB}(x) &=& B_j^+\sqrt{\frac{\hbar}{p(x)}}\exp\left(\int_{x_1+ (j-\frac{1}{2})a}^x \frac{p(y)}{\hbar}dy\right)     \cr
&&       + B_j^-\sqrt{\frac{\hbar}{p(x)}}\exp\left(-\int_{x_1+ (j-\frac{1}{2})a}^x \frac{p(y)}{\hbar}dy\right),
\end{eqnarray}
where $p(x)$ is defined as
\begin{equation}
p(x)= \sqrt{2m[V(x) -E]},
\end{equation}
with constant $ B_j^+$ and $B_j^-$.

In the forbidden region of quadratic potential near $x=x_1 +(j-1)a +l\sqrt{2\nu+1}$, we have
\begin{eqnarray}
&&\int_{x_1+ (j-\frac{1}{2})a}^x \frac{p(y)}{\hbar}dy\cr
&&~=-\int_{x_1+(j-1)a+l\sqrt{2\nu+1}}^{x_1+(j-\frac{1}{2})a}\frac{p(y)}{\hbar}dy \cr
&&~~~~~+\frac{1}{l}\int_{x_1 +(j-1)a +l\sqrt{2\nu+1}}^x \sqrt {\frac{(y-x_1-(j-1)a)^2}{l^2}-2\nu-1}dy\cr
&&~=-\int_{x_1+(j-1)a+l\sqrt{2\nu+1}}^{x_1+(j-\frac{1}{2})a}\frac{p(y)}{\hbar}dy
+\int_{\sqrt{2\nu+1}}^{\frac{z_{j-1}}{\sqrt 2}} \sqrt{u^2 -2\nu -1}du.
\end{eqnarray}
For $z_{j-1}\gg \sqrt{2\nu+1}$, we find that \cite{Furry}
\begin{eqnarray}
&&\int_{\sqrt{2\nu+1}}^{\frac{z_{j-1}}{\sqrt 2}} \sqrt{u^2 -2\nu -1}~du\cr
&&~= \frac{z_{j-1}^2}{4}-\frac{1}{2}\left( \nu +\frac{1}{2} \right)
-\left( \nu +\frac{1}{2} \right)\ln \frac{z_{j-1}}{\sqrt{\nu+\frac{1}{2}}} +O\left(\frac{ \nu +\frac{1}{2}}{z_{j-1}^2}\right)
\label{FurryA}
\end{eqnarray}
[we note the sign errors in Eqs.~(18) and (24) of Ref.~\cite{Song15} which are, however,  exclusively typographical and have no further effect], and, using the finite periodicity of $V(x)$,
we arrive at
\begin{equation}
\exp\left(\int_{x_1+ (j-\frac{1}{2})a}^x \frac{p(y)}{\hbar}dy\right)\simeq
\sqrt{\frac{\nu!g_\nu}{\sqrt{2\pi}}}\frac{e^{z_{j-1}^2/4}}{z_{j-1}^{1/2+\nu}}
\exp\left(-\int_{x_1+l\sqrt{2\nu+1}}^{x_1+\frac{a}{2}} \frac{p(y)}{\hbar}dy\right),~
\label{exp expansion}
\end{equation}
where
\begin{equation}
g_\nu =\frac{\sqrt{2\pi}}{\nu!}\left(\nu+\frac{1}{2}\right)^{\nu+\frac{1}{2}}e^{-\nu-\frac{1}{2}}.
\label{gnu33}
\end{equation}
In the quadratic region near $x=x_1 +(j-1)a +l\sqrt{2\nu+1}$ satisfying $z_{j-1}\gg \sqrt{2\nu+1}$, 
using $\frac{\hbar}{p(x)}\simeq \frac{\sqrt2 l}{z_{j-1}}$ and Eq.~(\ref{exp expansion}), we thus have
\begin{eqnarray}
\psi_j^{WKB}(x) &\simeq& 
B_j^+ \sqrt{\frac{\nu!g_\nu l}{\sqrt{\pi}}}\frac{e^{z_{j-1}^2/4}}{z_{j-1}^{\nu+1}}
\exp\left(-\int_{x_1+l\sqrt{2\nu+1}}^{x_1+\frac{a}{2}} \frac{p(y)}{\hbar}dy\right) \cr
&& + B_j^- \sqrt{\frac{l\sqrt{4\pi}}{\nu!g_\nu}}z_{j-1}^{\nu}e^{-z_{j-1}^2/4}
\exp\left(\int_{x_1+l\sqrt{2\nu+1}}^{x_1+\frac{a}{2}} \frac{p(y)}{\hbar}dy\right),
\label{LjWKB}
\end{eqnarray}
for $j=1,2,\ldots, N-1$. Introducing 
\begin{equation}
\epsilon_L^\nu= \sqrt{\frac{\nu!g_\nu }{\sqrt{\pi}}}\exp\left(-\int_{x_1+l\sqrt{2\nu+1}}^{x_1+\frac{a}{2}} \frac{p(y)}{\hbar}dy\right),
\label{epsilonL}
\end{equation}
we rewrite Eq.~(\ref{LjWKB}) as
\begin{equation}
\psi_j^{WKB}(x)\simeq B_j^+ \epsilon_L^\nu\sqrt{l}\frac{e^{z_{j-1}^2/4}}{z_{j-1}^{\nu+1}}+ B_j^-\sqrt{2l}
\frac{z_{j-1}^{\nu}e^{-z_{j-1}^2/4}}{\epsilon_L^\nu}.
\label{WKBzjggnu}
\end{equation}

In the quadratic region near  $x=x_1 +ja -l\sqrt{2\nu+1}$, for negative $z_{j}$ with $|z_{j}|\gg\sqrt{2\nu+1}$,
 we have
\begin{eqnarray}
&&\int_{x_1+(j-\frac{1}{2})a}^x \frac{p(y)}{\hbar}dy & \cr
&&=
 -\int_x^{x_1+ja-l\sqrt{2\nu+1}}\frac{p(y)}{\hbar}dy+\int_{x_1+(j-\frac{1}{2})a}^{x_1+ja-l\sqrt{2\nu+1}}\frac{p(y)}{\hbar}dy \cr
&&\simeq
-\frac{z_{j}^2}{4}+ (\nu+\frac{1}{2})\ln \frac{\sqrt{e}|z_{j}|}{\sqrt{\nu+\frac{1}{2}}}
 +\int_{x_1+\frac{1}{2}a}^{x_1+a-l\sqrt{2\nu+1}}\frac{p(y)}{\hbar}dy,
\label{eq26}
\end{eqnarray}
and thus
\begin{eqnarray}
\psi_j^{WKB}(x) &\simeq& 
B_j^+ \sqrt{\frac{l\sqrt{4\pi}}{\nu!g_\nu}}|z_{j}|^{\nu}e^{-z_{j}^2/4}
\exp\left(\int_{x_1+\frac{a}{2}}^{x_1+a-l\sqrt{2\nu+1}} \frac{p(y)}{\hbar}dy\right)   \cr
&& + B_j^- \sqrt{\frac{\nu!g_\nu l}{\sqrt{\pi}}}\frac{e^{z_{j}^2/4}}{|z_{j}|^{\nu+1}}
\exp\left(-\int_{x_1+\frac{a}{2}}^{x_1+a-l\sqrt{2\nu+1}} \frac{p(y)}{\hbar}dy\right) ,~~
\label{RjWKB}
\end{eqnarray}
for $j=1,2,\ldots, N-1$.
Introducing 
\begin{equation}
\epsilon_R^\nu= \sqrt{\frac{\nu!g_\nu }{\sqrt{\pi}}}\exp\left(-\int_{x_1+\frac{a}{2}}^{x_1+a-l\sqrt{2\nu+1}} \frac{p(y)}{\hbar}dy\right),
\label{epsilonR}
\end{equation}
we  rewrite Eq.~(\ref{RjWKB}) as
\begin{equation}
\psi_j^{WKB}(x)\sim  B_j^+\sqrt{2l} \frac{|z_{j}|^{\nu}e^{-z_{j}^2/4}}{\epsilon_R^\nu}
+ B_j^- \epsilon_R^\nu\sqrt{l}\frac{e^{z_{j}^2/4}}{|z_j|^{\nu+1}}.
\label{mWKBzjggnu}
\end{equation}

\section{Relations for the coefficients: A quantization condition }
\label{sec:qcondition}
As $\psi(x)$ is described by  $\psi_j(x)$ or by $\psi_j^{WKB}(x)$ depending on the regions, if  $\psi(x)$ is  described by two different functions in an overlapping region  (see Fig.~\ref{figure2}),  the functions should match onto each other for the continuity in the region, which gives the relations for the coefficients introduced in applying Dekker's method. Since we are interested in the low-lying states with $a\gg l$, we assume  that 
\begin{equation}
\nu= n+\delta_n,
\label{nudelta}
\end{equation}
where $\delta_n\ll 1$ and  $n$ is a small non-negative integer.

Comparing the asymptotic expansion of $\psi_0(x)$ given by Eq.~(\ref{z0ggnu}) and that of $\psi_1^{WKB}(x)$ given in Eq.~(\ref{WKBzjggnu}) for $j=1$ in the overlapping region near $x=x_1+l\sqrt{2n+1}$, we have
\begin{eqnarray}
&&(-1)^n  C_0=\frac{\sqrt{2l}}{\epsilon_L^n}B_1^-, \label{0pD}\\
&&(-1)^{n+1}n!\sqrt{2\pi}\delta_n C_{0}=\sqrt{l}\epsilon_L^nB_1^+. \label{0pH}
\end{eqnarray}
For $j=2,3,\ldots,N-1$, in the region of quadratic potential near $x=x_1+(j-1)a+l\sqrt{2n+1}$, if $z_{j-1} \gg\sqrt{2n+1}$, the wave function is described by $\psi_{j-1}(x)$ as well as
$\psi_j^{WKB}(x)$; then, the asymptotic relations in Eqs.~(\ref{zjggnu}) and (\ref{WKBzjggnu}) imply that
\begin{eqnarray}
&&C_{j-1}^+ + (-1)^n C_{j-1}^-  =\frac{\sqrt{2l}}{\epsilon_L^n}B_j^- ,\label{jpH}\\
&&(-1)^{n+1} n!\sqrt{2\pi}\delta_n C_{j-1}^-= \sqrt{l}\epsilon_L^nB_j^+.\label{jpD}
\end{eqnarray}
For $j=1,2,\ldots,N-2$, 
in the region of quadratic potential near $x=x_1+ja-l\sqrt{2n+1}$, if $-z_{j} \gg\sqrt{2n+1}$, the wave function is described by $\psi_{j}(x)$ as well as
$\psi_j^{WKB}(x)$; then, the asymptotic relations in Eqs.~(\ref{mzjggnu}) and (\ref{mWKBzjggnu}) imply that
\begin{eqnarray}
&& (-1)^n C_{j}^+ + C_{j}^-  =\frac{\sqrt{2l}}{\epsilon_R^n}B_j^+,\label{jmH}\\
&&(-1)^{n+1} n!\sqrt{2\pi}\delta_n C_{j}^+= \sqrt{l}\epsilon_R^n B_j^-.\label{jmD}
\end{eqnarray}
Comparing the asymptotic expansion of $\psi_{N-1}(x)$ of Eq.~(\ref{zN-1ggnu}) and that of $\psi_{N-1}^{WKB}(x)$ in the overlapping region near $x=x_1+(N-1)a-l\sqrt{2n+1}$, we have
\begin{eqnarray}
&&(-1)^n C_{N-1}=\frac{\sqrt{2l}}{\epsilon_R^n}B_{N-1}^+, \label{N-1pH}\\
&&(-1)^{n+1}n!\sqrt{2\pi}\delta_n C_{N-1}=\sqrt{l}\epsilon_R^nB_{N-1}^-. \label{N-1pD}
\end{eqnarray}

While we have introduced $4(N-1)$ coefficients:  $ C_0$, $B_j^+$, $B_j^-$ ($j=1,2,\ldots, N-1$), $C_k^+$, $C_k^-$ ($k=1,2,\ldots, N-2$),  $C_{N-1}$, Eqs.~(\ref{0pD}-\ref{N-1pD}) constitute  $4(N-1)$ linear homogeneous equations for the coefficients. Instead of analyzing these equations directly, we wish to extract $N$ equations for the $N$ coefficients:  $ C_0$, $ C_j^++(-1)^n C_j^-~(j=1,2,\ldots, N-2)$,  $C_{N-1}$.

Substituting $j+1$ for $j$ in Eqs.~(\ref{jpH},\ref{jpD},\ref{jmH}), we have
\begin{eqnarray}
&&C_{j}^+ + (-1)^n C_{j}^-  =\frac{\sqrt{2l}}{\epsilon_L^n}B_{j+1}^- ,\label{j+1pH}\\
&&(-1)^{n+1} n!\sqrt{2\pi}\delta_n C_{j}^-= \sqrt{l}\epsilon_L^nB_{j+1}^+ ,\label{j+1pD}\\
&& (-1)^n C_{j+1}^+ + C_{j+1}^-  =\frac{\sqrt{2l}}{\epsilon_R^n}B_{j+1}^+,\label{j+1mH},
\end{eqnarray}
respectively. We rewrite Eq.~(\ref{j+1pD}) as
\begin{equation}
- n!\sqrt{2\pi}\delta_n[C_j^+ +(-1)^{n} C_{j}^-]+ n!\sqrt{2\pi}\delta_nC_j^+= \sqrt{l}\epsilon_L^nB_{j+1}^+.
\label{modifiedf}
\end{equation}
Using Eqs.~(\ref{jmD}), (\ref{jpH}), and (\ref{j+1mH}), we then arrive at
\begin{eqnarray}
&&\delta_n(C_j^+ +(-1)^{n}C_{j}^-)\cr
&&+\frac{(-1)^n\epsilon_L^n\epsilon_R^n}{2\sqrt{\pi}n!}
\left[(C_{j-1}^++(-1)^nC_{j-1}^-)+(C_{j+1}^++(-1)^nC_{j+1}^-)\right]=0  \label{Cj}
\end{eqnarray}
which is valid for $j=2,3,\ldots,N-3$.

Equation (\ref{jmH}) for $j=1$
and Eq.~(\ref{0pH}) yield
\begin{equation}
\delta_n  C_0 +\frac{\epsilon_L^n\epsilon_R^n}{2\sqrt{\pi}n!}\left( C_{1}^++(-1)^nC_{1}^-  \right) =0.\label{C0rel}
\end{equation}
Using Eq.~(\ref{jmD}) for $j=1$ and Eq.~(\ref{0pD}),  we have
\begin{equation}
 n!\sqrt{2\pi}\delta_n  C_1^+=-\frac{\epsilon_L^n\epsilon_R^n}{\sqrt2}C_0.
\label{C1C0}
\end{equation}
Plugging the expression of $C_1^+$ in Eq.~(\ref{C1C0})  into Eq.~(\ref{modifiedf}) for $j=1$, and using Eq.~(\ref{jmD}), we find
\begin{equation}
\delta_n(C_1^+ +(-1)^{n}C_{1}^-)
+\frac{\epsilon_L^n\epsilon_R^n}{2\sqrt{\pi}n!}
\left[ C_0+(-1)^n (C_{2}^++(-1)^nC_{2}^-)\right]=0.\label{C1rel}
\end{equation}

Using Eq.~(\ref{N-1pD}) and Eq.~(\ref{jpH}) with $j=N-1$, we have
\begin{equation}
\delta_n C_{N-1} +(-1)^{n}\frac{\epsilon_L^n\epsilon_R^n}{2\sqrt{\pi}n!}\left( C_{N-2}^++(-1)^nC_{N-2}^-  \right) =0.\label{CN-1rel}
\end{equation}
By a  similar procedure to the previous cases, Eq.~({\ref{N-1pH}) can be used to give
\begin{eqnarray}
&&\delta_n(C_{N-2}^+ +(-1)^{n}C_{N-2}^-)\cr
&&+\frac{(-1)^n\epsilon_L^n\epsilon_R^n}{2\sqrt{\pi}n!}
\left[(C_{N-3}^++(-1)^nC_{N-3}^-)+C_{N-1}\right]=0 . \label{CN-2}
\end{eqnarray}

 Eqs.~(\ref{C0rel}), (\ref{C1rel}), (\ref{Cj}),  (\ref{CN-2}), (\ref{CN-1rel}) constitute the $N$ equations. 
By defining $\tilde C_0=(-1)^n  C_0$, $C_j=C_j^+ +(-1)^{n}C_{j}^-$ $(j=1,2,\ldots,N-2)$, 
\begin{equation}
{\Delta}_n ={\hbar\omega}\frac{\epsilon_L^n\epsilon_R^n}{\sqrt{\pi}n!}
=g_n \frac{\hbar\omega}{\pi}\exp\left(-\int_{x_1+l\sqrt{2n+1}}^{x_1+a-l\sqrt{2n+1}}
\frac{p(y)}{\hbar} dy \right),
\label{Dnfirst}
\end{equation}
and a $N\times N$ symmetric Toeplitz tridiagonal matrix
\begin{equation}
{\bf T}=
\left( \begin{array}{cccccccc}
0   &1    &0    & 0   & 0    &0     & \ldots     &0 \cr
1   &0    &1    & 0   & 0    &0     & \ldots     &0 \cr
0   &1    &0    & 1   & 0    &0     & \ldots     &0 \cr
0   &0    &1    & 0   & 1    &0     & \ldots     &0 \cr
\vdots&\ddots&\ddots&\ddots&\ddots&\ddots&\ddots&\vdots  \cr
0   &\ldots&0   &0    &1    & 0   & 1    &0  \cr
0   &\ldots&0   &0    &0    &1    & 0    & 1    \cr
0   &\ldots&0   &0    &0    &0    & 1    & 0 
\end{array}\right),
\end{equation}
the $N$ equations can be written as
\begin{equation}
\left(\hbar\omega {\delta}_n{\bf I} +\frac{(-1)^n{\Delta}_n}{2}{\bf T} \right)
\left( \begin{array}{c}
\tilde{C}_0\cr
C_1\cr
C_2\cr
\vdots\cr
C_{N-1}\end{array}\right)=0,
\end{equation}
where ${\bf I}$ denotes unit matrix. 

If $ (C_0,C_1,\ldots, C_{N-1})=(0,0,\ldots,0)$, for ${\delta}_n\neq  0$,  Eqs.~(\ref{0pD}-\ref{N-1pD}) show that all the $4(N-1)$ coefficients are zero, implying the wave function $\psi(x)$ vanishes everywhere. Hence, for an acceptable wave function, $\delta_n$ should satisfy the quantization condition:
\begin{equation}
\det\left(\hbar\omega {\delta}_n{\bf I} +\frac{(-1)^n{\Delta}_n}{2}{\bf T} \right)=0.
\end{equation}
As is well-known (see, e.g., Ref.~\cite{circulant,Sacchetti}), the $N$ eigenvalues of ${\bf T}$ are $2\cos\frac{s\pi}{N+1}$ ($s=1,2,\ldots, N$). 
Thus the quantization condition is satisfied when $\hbar\omega {\delta}_n=(-1)^{n+1}\Delta_n\cos\frac{s\pi}{N+1}$ which, with Eqs.~(\ref{Enu2}) and (\ref{nudelta}), implies that the energy eigenvalue of the multiple-well system are given as
\begin{equation}
E_n(s)= E_n^{(0)}+(-1)^{n+1}\Delta_n\cos\frac{s\pi}{N+1};~~s=1,2,\ldots,N,
\label{Ens}
\end{equation}
when
\begin{equation}
(\tilde C_0,C_1,\ldots, C_{N-1})= \left(\sin\frac{s\pi}{N+1},\sin\frac{2s\pi}{N+1},\ldots, ,\sin\frac{Ns\pi}{N+1}\right),
\label{eigenCoeff}
\end{equation}
where $E_n^{(0)}$ is the energy eigenvalue of the corresponding harmonic oscillator: $E_n^{(0)}=V_0+(n+\frac{1}{2})\hbar\omega$.

The cases of $\delta_n=0$ appear in Eq.~(\ref{Ens}) when $N$ is odd and $s=(N+1)/2$; in these case,  Eqs.~(\ref{0pH},\ref{jpD},\ref{jmD},\ref{N-1pD}) show $B_j^+=B_j^-=0$ $(j=1,2,\ldots,N-1)$, and then Eqs.~(\ref{0pD},\ref{jpH},\ref{N-1pH}) imply $ (C_0,C_1,\ldots, C_{N-1})=(0,0,\ldots,0)$ to denote  that the validity of Dekker's method may not be provided in these cases (see the next section). 
Indeed, it has been discussed for $N=2$ that  two-level approximation would be appropriate  in the limit of $\delta_n\rightarrow 0$ \cite{Song15}, and thus $N$-level approximation will be explored in the next section.

\section{ $N$-level approximation}
\label{sec:N-level}
 If barrier penetration could be ignored in the multiple-well potential, for the low-lying states, we may have $N$-fold degenerate states which are either the individual states localized  in each  separate well  or any combinations of them. Barrier penetration lifts the degeneracy, to select  specific linear combinations as the eigenstates. 

In an extension of the two-level approximation for a double-well potential, for non-negative integer $n$, we hypothesize an  approximate real solution  $\tilde{\psi}_n (j;x)~(\equiv <x|\tilde\psi_n(j)>)$  to the Schr\"{o}dinger equation which is primarily  localized in the classically allowed region around a minimum at $x= x_1+ja$ with energy $E_n^{(0)}$, and has small probability distribution in the two classically forbidden regions attached to it, with vanishing amplitude for $x\geq x_1+(j+1)a-l\sqrt{2n+1}$ and for $x \leq x_1+(j-1)a+l\sqrt{2n+1}$.  Then,  in the quadratic region containing $x= x_1+ja$, $\tilde{\psi}_n (j;x)$  is naturally approximated by the wave function of the $n$th excited state of harmonic oscillator centered at $x=x_1+ja$:
 \begin{eqnarray}
&&\psi_{n}^{sho}(j;x)=\frac{1}{\pi^\frac{1}{4}\sqrt{2^{n}n!l}}
H_{n}\left(\frac{x-x_1-aj}{l}\right)e^{-{(x-x_1-aj)^2}/{2l^2}}\nonumber\\
&&~=\frac{1}{\pi^\frac{1}{4}\sqrt{n!l}}z_j^n\left[1-\frac{n(n-1)}{2z_j^2}+\cdots\right]e^{-z_j^2/4}, 
\label{sho}
\end{eqnarray}
where  $H_{n}$ denotes the $n$th order  Hermite polynomial.

In the forbidden region attached to the left-hand side of the quadratic region containing $x=x_1+ja$,  the pertinent WKB  approximation to  $\tilde{\psi}_n (j;x)$ is 
\begin{equation}
\psi_{WKB}^L(j;x)= N_{jL} ~\sqrt{\frac{\hbar}{p(x)}}e^{\int_{x_1+(j-\frac{1}{2})a}^x \frac{p(y)}{\hbar} dy};~~
j=1,2,\ldots, N-1
\label{WKBjL}
\end{equation}
the amplitude of which increases as $x$ increases, where $ N_{jL}$ is a constant.
In the region of quadratic potential near $x=x_1+ja-l\sqrt{2n+1}$ satisfying $z_j\ll -\sqrt{2n+1}$,  using Eq.~(\ref{eq26}) and  $\frac{\hbar}{p(x)}\simeq \frac{\sqrt2 l}{|z_{j}|}$, we  find
\begin{equation}
\psi_{WKB}^L(j;x)\simeq N_{jL} \sqrt{\frac{2l\sqrt\pi}{n!g_n}}|z_j|^ne^{-z_j^2/4+\int_{x_1+\frac{1}{2}a}^{x_1+a-l\sqrt{2n+1}} \frac{p(y)}{\hbar} dy}.
\end{equation}
As $\psi_{WKB}^L(j;x)$  and  $\psi_{n}^{sho}(j;x)$ are approximations for the same wave function,  $\psi_{WKB}^L(j;x)$ should match on to $\psi_{n}^{sho}(j;x)$ in the overlapping region of  $z_j\ll -\sqrt{2n+1}$: Comparing the leading terms in the region, we thus have
\begin{equation}
N_{jL}=(-1)^n\sqrt\frac{g_n}{2\pi} \frac{1}{l}\exp\left(-\int_{x_1+\frac{a}{2}}^{x_1+a-l\sqrt{2n+1}}\frac{p(y)}{\hbar}dy\right)
\equiv N_{L},
\label{NjLNL}
\end{equation}
where the  constant $N_{L}$ is introduced since $N_{jL}$ does not depend on $j$.

In the forbidden region attached to the right-hand side of the well at  $x= x_1+ja$, the pertinent WKB wave function, with constant $ N_{jR}$, is 
\begin{equation}
\psi_{WKB}^R(j;x)= N_{jR} ~\sqrt{\frac{\hbar}{p(x)}}e^{\int_x^{x_1+(j+\frac{1}{2})a} \frac{p(y)}{\hbar} dy};~~
j=0,1,\ldots N-2
\end{equation}
the amplitude of which decreases as $x$ increases. Similarly,
in the quadratic and forbidden region satisfying $z_j\gg \sqrt{2n+1}$, we have
\begin{equation}
\psi_{WKB}^R(j;x)\simeq N_{jR} \sqrt{\frac{2l\sqrt\pi}{n!g_n}}z_j^n e^{-z_j^2/4+\int_{x_1+l\sqrt{2n+1}}^{x_1+\frac{1}{2}a} \frac{p(y)}{\hbar} dy}.
\end{equation}
Comparing the leading term  of $\psi_{WKB}^R(j;x)$ and that of $\psi_{n}^{sho}(j;x)$, we  find 
\begin{equation}
N_{jR}=\sqrt\frac{g_n}{2\pi} \frac{1}{l}\exp\left(-\int_{x_1+l\sqrt{2n+1}}^{x_1+\frac{a}{2}} \frac{p(y)}{\hbar}dy\right)
\equiv N_R,
\label{N_jR}
\end{equation}
where, again, $N_R$ is a constant which does not depend on $j$.
Indeed, the finite periodicity implies
\begin{equation}
\tilde{\psi}_n (j;x)=\tilde{\psi}_n (1;x-(j-1)a)
\label{finite periodicity} 
\end{equation} 
for $j=1,2,\ldots,N-2$, as can be seen through the explicit constructions.

As an  approximation to include  the tunneling  effect, we may restrict our attention on the $N$-dimensional subspace spanned  by $\{ |\tilde{\psi}_n (j)>;~ j=0,1, \ldots, N-1\}$, as   all the states have the approximate energy eigenvalue $E_n^{(0)}$.  Since, for $|j-k|\geq 2$, $\tilde{\psi}_n (j;x)$ does not overlap with   $\tilde{\psi}_n (k;x)$ by our construction, we have the matrix element of the Hamiltonian 
\begin{equation}
<\tilde\psi_n(k)|H|\tilde\psi_n(j)>=\int_{-\infty}^\infty \tilde{\psi}_n^* (k;x) H \tilde{\psi}_n (j;x) =0~~~({\rm for} ~~|j-k|\geq 2).
\label{2-off}
\end{equation} 
As we are considering real  $\tilde{\psi}_n (j;x)$ we also have 
\[<\tilde\psi_n(j+1)|H|\tilde\psi_n(j)>=<\tilde\psi_n(j)|H|\tilde\psi_n(j+1)>.\]

 As a further approximation for estimating  the matrix element of $<\tilde\psi_n(j+1)|H|\tilde\psi_n(j)>$,  we restrict our attention on the two-dimensional subspace spanned by 
$|\tilde{\psi}_n (j)>$ and $|\tilde{\psi}_n (j+1)>$, in which 
$\tilde\psi_n^\pm(j;x)=\frac{1}{\sqrt{2}}((\tilde\psi_n(j;x)\mp(-1)^n\tilde\psi_n(j+1;x))$ satisfy the Schr\"{o}dinger equation:
\begin{equation}
H\tilde\psi_n^\pm(j;x)\simeq  E_n^\pm \tilde\psi_n^\pm(j;x) 
\label{Seqjpm}
\end{equation}
with the eigenvalues $ E_n^\pm= E_0(n)\pm \frac{\tilde{\Delta}_n(j)}{2}$ (see, e.g., Refs.~\cite{DekkerPRA,Song}), where
\begin{equation}
\tilde{\Delta}_n(j)= 2\times (-1)^{n+1}<\tilde\psi_n(j+1)|H|\tilde\psi_n(j)>.
\label{delta_napp}
\end{equation}
From the definitions of the localized wave functions,  we have
\begin{eqnarray}
&&\int_{x_1+(j+\frac{1}{2})a}^\infty \tilde\psi_n^2(j+1;x) \approx 1,~~~~~
\int_{x_1+(j+\frac{1}{2})a}^\infty \tilde\psi_n^2(j;x) \approx 0 ,\cr
&&
\int_{x_1+(j+\frac{1}{2})a}^\infty \tilde\psi_n(j;x))\tilde\psi_n(j+1;x) \approx 0.
\label{Nlevelnorm}
\end{eqnarray}
We multiply the Schr\"{o}dinger equation (\ref{Seqjpm}) for $\tilde\psi_n^+(j;x)$ by $\tilde\psi_n^-(j;x)$, and  the equation for 
$\tilde\psi_n^-(j;x)$ by $\tilde\psi_n^+(j;x)$. Similarly as in 
Ref.~\cite{DekkerPRA}, subtracting the two resulting expressions and integrating over $x\in [x_1+(j+\frac{1}{2})a,\infty)$,  we arrive at
\begin{eqnarray}
&&\tilde{\Delta}_n(j) \cr
&&~\simeq  (-1)^{n}\frac{\hbar^2}{m}
\left. \left(
\tilde\psi_n(j;x)\frac{d \tilde\psi_n(j+1;x)}{dx} -   \tilde\psi_n(j+1;x)\frac{d\tilde \psi_n(j;x)}{dx}
\right)\right|_{x=x_1+(j+\frac{1}{2})a}  \cr
& &~= (-1)^{n}\frac{2\hbar^2}{m}N_{j+1L}N_{jR}  = (-1)^{n}\frac{2\hbar^2}{m}N_LN_R   \cr
&&~= \Delta_n.
\label{twoDelta}
\end{eqnarray}

{\em On the assumption that} we could neglect the mutual overlap of the wave functions so that
\begin{equation}
<\tilde\psi_n(j+1)|\tilde\psi_n(j)>\simeq0,
\label{SBA}
\end{equation}
 the $N\times N$ Hamiltonian matrix in the subspace spanned by  $|\tilde\psi_n(j)> ~(j=0,1,2,\ldots,N-1)$
is written as
\begin{equation}
H=\left(E_n^{(0)}{\bf I} +\frac{(-1)^{n+1}{\Delta}_n}{2}{\bf T} \right),
\end{equation}
whose eigenvalues are $E_n(s)$ given by Eq.~(\ref{Ens}).
The wave function corresponding to the eigenvalue $E_n(s)$ is given as [see Eq.~(\ref{eigenCoeff})]
\begin{equation}
\tilde{\psi}_{E_n(s)}(x)=\sqrt{\frac{2}{N+1}}\sum_{j=0}^{N-1}\tilde\psi_n(j;x)\sin\frac{(j+1)s\pi}{N+1},
\label{waveNlevel}
\end{equation}
where the normalization constant is determined from the fact that \cite{Sacchetti}
\[\sum_{j=0}^{N-1} \sin^2 \frac{(j+1)s\pi}{N+1} =\frac{N+1}{2}.\]
Indeed, the fact 
\[
D_{n}(z)= 2^{-\frac{n}{2}}e^{-\frac{z^2}{4}}H_{n}\left( \frac{z}{\sqrt2}\right)
\]
implies   $\psi_0(x) \rightarrow \pi^{\frac{1}{4}}\sqrt{n!l}\tilde C_0\psi_n^{sho}(0;x)$ and 
$\psi_j(x) \rightarrow\pi^{\frac{1}{4}}\sqrt{n!l}C_j\psi_n^{sho}(j;x)$  $(j=1,2,\ldots, N-1)$
in the limit as $\nu$ goes to $n$, to show  that the wave function given in the  $N$-level approach is equivalent to that found through Dekker's method.

\section{ Comparison and applications }
\label{sec:comparison}
For the $(N=2)$ double-well  potential, Eq.~(\ref{Ens}) shows that the energy eigenvalues associated with $E_n^{(0)}$ are $E_n^{(0)}\pm \frac{\Delta_n}{2}$, with the level splitting $\Delta_n$ which is in agreement with the expression in Refs.~\cite{DekkerPRA,Song}.
For large $N$, the eigenvalues associated with $E_n^{(0)}$ form bands, and the widths of the energy bands become $2 \Delta_n$  as $N$ goes to infinity.

\subsection{Comparison with the strong bonding approximation}
\label{sec:TBA}
For a periodic  potential $V_p(x)$ which is equal to  $V(x)$ for  $x_1 \leq  x\leq  x_1 +(N- 1)a$,  but with the full periodicity $V_p(x+a)= V_p(x)$ for all $x$, we construct the localized approximate eigenfunctions by considering that Eq.~(\ref{finite periodicity}) is true for all integer $j$. The (unnormalized) wave function 
\begin{equation}
\varphi_{k,n} (x)= \sum_{j=-\infty}^\infty e^{ikaj}\tilde{\psi}_n(j;x),
\end{equation}
then satisfies the Bloch condition: $\varphi_{k,n} (x+a)=e^{ika}\varphi_{k,n} (x)$,
with  the Bloch wavenumber $k$ satisfying
\begin{equation}
-\frac{\pi}{a} \leq k <\frac{\pi}{a}.
\label{Brillouin}
\end{equation} 
If we assume Eq.~(\ref{SBA}) is valid  for all  integer $j$, as in \cite{CDL}, the wave function  $\varphi_{k,n} (x)$ can be  shown to be an eigenfunction of the periodic system with the energy eigenvalue:
\begin{equation}
E_{n}^p(k)= E_n^{(0)}+(-1)^{n+1}\Delta_n\cos ka.
\label{Eperiodic}
\end{equation}
Equation (\ref{Eperiodic}) suggests that the strong bonding approximation of  Ref.~\cite{CDL} is, in fact, related to the rigorous tight-binding approach, and  $E_{n}^p(k)- E_n^{(0)}$  may correspond to the wave-number dependent corrections to the energy expectation value contributed by the nearest neighbors in the tight-binding approximation  of one dimension \cite{Harrison,AM}.

When the Bloch phase, $ka$, is equal to $s\pi/(N+1)$, $E_{n}^p(k)$ coincides with the  eigenvalue $E_n(s)$ of the finite periodic system given in Eq.~(\ref{Ens}). 
For the system of a finite periodic potential which consists  of $N$ square wells, through the transfer matrix method, it has been shown that the low-lying states are described by the Bloch phases  $s\pi/(N+1)$   if the single square well is deep and wide \cite{SSWM}, while the WKB approximation may not be  useful for  rectangular potential curves (see, e.g., Ref.~\cite{Furry}).

In spite of the similarity between the finite periodic system and fully periodic system, we note the differences:  The eigenfunctions of the finite system do not fulfill the periodicity property  $|\tilde{\psi}_{E_n(s)}(x+a)|^2=|\tilde{\psi}_{E_n(s)}(x)|^2$ in the domain of the periodicity of $V(x)$ as detailed in Ref.~\cite{Pereyra}, and there is no degeneracy of the eigenvalues $E_n(s)$ in the finite system while $E_{n}^p(k) =E_{n}^p(-k)$. In \ref{sec:appendix}, we analyze the lowest band of a two-dimensional system in which  most of the energy eigenvalues are degenerate in the large-$N$ limit and  the eigenfunctions fulfill the periodicity property under the discrete rotations.

\subsection{The cosine potential}
In order to compare the result with the  rigorous expression for the widths of the low-lying energy bands of the  Mathieu equation,  we consider the $2N$-well potential
\begin{equation}
V_c(x)=2q \cos \frac{2x}{l_c}  ~~~{\rm for} ~~|x|<N\pi l_c,
\end{equation}
with positive constant $q$ and $l_c$,  assuming that $V_c(x)$ is  monotonically decreasing (increasing) for $x\leq -N\pi l_c$ ($x\geq N\pi l_c$).
We find the approximate expression for $\omega$ and $l$ as:
\begin{eqnarray}
&&\omega=\left( \left.\frac{1}{m}\frac{d^2V_c(x)}{dx^2}\right|_{x=\frac{\pi}{2} l_c}
 \right)^{1/2}=\frac{1}{l_c}\sqrt{\frac{8q}{m}},
\label{cos omega}\\
&&l=\sqrt{{\hbar}\over{m\omega}}=\frac{\sqrt{\hbar l_c}}{(8mq)^{1/4}}.\nonumber
\end{eqnarray}
Introducing 
\begin{equation}
\varphi_M= \frac{\pi}{2}- \frac{l}{l_c}\sqrt{2n+1}
= \frac{\pi}{2}- \sqrt{(n+\frac{1}{2})\sqrt{\frac{\hbar^2}{2ml_c^2}\frac{1}{q}}},
\end{equation}
 for the energy $E_n^{(0)}$ [$=-2q+\hbar\omega(n+\frac{1}{2}$)], it may be appropriate to take $x=\pm l\varphi_M$ as  the turning points adjacent to $x=0$, as we are interested in in the limit of $\varphi_M \rightarrow \frac{\pi}{2}$.

The integral in the exponential of Eq.~(\ref{Dnfirst}) can thus be approximated as 
\begin{eqnarray}
&&\int_{x_1+l\sqrt{2n+1}}^{x_1+a-l\sqrt{2n+1}}\frac{p(y)}{\hbar} dy \simeq
\int_{-l_c \varphi_M}^{l_c \varphi_M}\frac{p(y)}{\hbar} dy \cr
&&\simeq \frac{2l_c\sqrt{mq}}{\hbar}\int_{-\varphi_M}^{\varphi_M}\sqrt{\cos(2\varphi)-\cos(2\varphi_M)} d\varphi  \cr
&&=\frac{4l_c\sqrt{2mq}}{\hbar}\left[ E(\sin \varphi_M )-\cos^2\varphi_M K(\sin \varphi_M )\right]
\cr
&&\simeq \frac{4l_c\sqrt{2mq}}{\hbar}-(n+\frac{1}{2}) -(n+\frac{1}{2})\ln\frac{16l_c\sqrt{2mq}}{\hbar(n+\frac{1}{2})}
\label{cos integral}
\end{eqnarray}
in the limit  of $\varphi_M \rightarrow \frac{\pi}{2}$, where $E$ and $K$ denote the complete elliptic integrals. 
Substituting Eqs.~(\ref{gnu33},\ref{cos omega},\ref{cos integral}) into Eq.~(\ref{Dnfirst}), we find the widths of the narrow bands:
\begin{equation}
2{\Delta}_n\simeq \frac{\hbar^2}{2ml_c^2}\frac{ 2^{4n+5}}{n!}\sqrt{\frac{2}{\pi}}
\left(\frac{2ml_c^2 }{\hbar^2}q\right)^{\frac{n}{2}+\frac{3}{4}}\exp\left(-4\sqrt{\frac{2ml_c^2}{\hbar^2}q}\right),
\end{equation}
which, if we take $\frac{\hbar^2}{2ml_c^2 }=1$, reproduces the known result (see, e.g.,  Ref.~\cite{Koch,CSDG,nist,CUMS}) at the leading order.

\section{ Concluding remarks}
\label{sec:concluding}
	We have analyzed the system of the finite periodic multiple-well potential using Dekker's method, to find a formula for the energy eigenvalues of the low-lying bands which could reproduce the rigorous mathematical expression for the widths of the narrow energy bands of  the Mathieu equation; in this method,  the assumption that the wells are parabolic with $a\gg l$  is made, and then wave function matching determines the formula. The same formula has also been derived through the $N$-level approximation. Though  more calculations are involved in applying Dekker's method, in the derivation of the formula  through the $N$-level approach, in addition to the assumption of  the parabolicity  used for constructing and normalizing $\tilde{\psi}_n(j;x)$, {\em other assumptions or approximations} such as those in Eqs.~(\ref{2-off}-\ref{delta_napp}, \ref{SBA})   have  {\em additionally} been  made.

The energy eigenvalues for the fully periodic potential which coincides with the finite periodic multiple-well potential on a finite domain have also been explicitly written in terms of the potential within the strong bonding approximation, and it is found that the eigenvalues of the $N$-well potential are  those that the eigenvalue formula of the fully periodic system gives  at some discrete Bloch wavenumbers (phases). While the  discrete Bloch  phases of our system have already been noticed in a related problem \cite{SSWM}, it has been known that different sets of Bloch wavenumbers  be used depending on the boundary conditions \cite{PP}, which  imply that our result would be valid within the boundary condition prescribed in Section \ref{sec:wave}. Specifically, the formula will be valid upon a boundary condition which is compatible with $\delta_n\ll1$ [see Eq.~(\ref{nudelta})]; for an example, if infinite wall is located near $x=x_1$ (or $x=x_1+(N-1)a$), the formula given here may not be valid.

The $N$-well system in the large-$N$ limit is different from the fully periodic system, as the  discrete Bloch wavenumbers of the  $N$-well system $\frac{s\pi}{a(N+1)}$ ($s=1,2,\ldots, N$) in the limit densely fill the the interval $[0,\frac{\pi}{a}]$ which is just the half of the first Brillouin zone given in Eq.~(\ref{Brillouin}) and  the eigenvalues are not degenerate; in this respect, a two-dimensional model is analyzed in the \ref{sec:appendix}. In spite of that eigenfunctions of the one-dimensional $N$-well system do not fulfill the periodicity property, if we ignore the overlaps between $\tilde\psi_n(j;x)$ and  $\tilde\psi_n(j+1;x)$, using   Eq.~(\ref{waveNlevel}) and $(-1)^j\sin \frac{(j+1)(N+1-s)\pi}{N+1}=\sin \frac{(j+1)s\pi}{N+1}$, we find
$|\tilde{\psi}_{E_n(N+1-s)}(x)|^2\simeq|\tilde{\psi}_{E_n(s)}(x)|^2$ which may be the {\em intraband symmetry} 
found  numerically in the related problems \cite{Pereyra}.

 \appendix
\section{A two-dimensional  multiple-well potential  }
\label{sec:appendix}
Let $v(x,y)$ be a smooth function on $R^2$, invariant under continuous rotation about the origin $(0,0)$.  We also assume that $v(x,y)$ has a quadratic minimum at
the origin and the potential is written in the quadratic region around the origin as
\begin{equation}
v(x,y)= V_0+ \frac{m\omega^2}{2}\rho^2,
\end{equation}
with $ \rho=\sqrt{x^2+y^2}$ and $-V_0 \gg \hbar\omega$. The wave function 
\[
\phi(x,y)=\frac{1}{l\sqrt{\pi}}e^{-\rho^2/2l^2}
\]
is then an approximate solution to the Schr\"{o}dinger equation for  $v(x,y)$.
Further assuming that  $v(x,y) =0$ for $ \rho>  r \sin\frac{\pi}{N}$ with a positive constant $r$, we consider the $N$-well potentials of the kind
\begin{equation}
V_{2d}(x,y)= \sum_{j=0}^{N-1} v(x-r\cos\frac{2j\pi}{N}, y-r\sin\frac{2j\pi}{N}).
\end{equation}

A coordinate rotation of an integer $k$
\begin{equation}
x_k'=x\cos\frac{2k\pi}{N}+y\sin\frac{2k\pi}{N},~~~~y_k'=-x\sin\frac{2k\pi}{N}+y\cos\frac{2k\pi}{N}
\end{equation}
gives the relation:
\begin{eqnarray}
&&(x-r\cos\frac{2j\pi}{N})^2+(y-r\sin\frac{2j\pi}{N})^2\cr
&&=(x_k'-r\cos\frac{2(j-k)\pi}{N})^2+(y_k'-r\sin\frac{2(j-k)\pi}{N})^2.
\end{eqnarray}
Introducing 
\begin{equation}
\phi_j(x,y)= \phi( x-r\cos\frac{2j\pi}{N}, y-r\sin\frac{2j\pi}{N} );~~j=0,1,\ldots,N-1,
\end{equation}
we then find
\begin{equation}
V_{2d}(x,y)=V_{2d}(x_k',y_k'),
\end{equation}
\begin{equation}
\phi_j(x,y)=\phi_{(j-k)~ {\rm mod}~ N}(x_k',y_k'),~~\phi_j(x_k',y_k')=\phi_{(j+k)~ {\rm mod}~ N}(x,y),
\label{DRotation}
\end{equation}
where $n~{\rm mod}~ N= n-N\floor*{n}.$ 

Defining
\begin{equation}
h_j= <\phi_0|H_{2d}|\phi_j>;~~j=0,1,\ldots,N-1,
\end{equation}
with $H_{2d}=-\frac{\hbar^2}{2m}\left( \frac{\partial^2}{\partial x^2}+\frac{\partial^2}{\partial y^2}\right) +V(x,y)$, and using 
\begin{equation}
 <\phi_k|H_{2d}|\phi_j>=h_{(j-k)~mod~N},
\end{equation}
on the assumption that  $<\phi_k|\phi_j>=\delta_{kj}$, the Hamiltonian  in the subspace spanned by $\{|\phi_j>|j=0,1,2,\ldots,N-1\}$ is given as a circulant matrix
\begin{equation}
{H_{2d}}=
\left( \begin{array}{cccccccc}
h_0   &h_1    &h_2           & \ldots     &h_{N-1} \cr
h_{N-1}   &h_0   &h_1      &     & h_{N-2}      \cr
\vdots    &h_{N-1}   &h_0       & \ddots     &\vdots \cr
h_2  & &\ddots&\ddots     &h_1 \cr
h_1   &\ldots &       & h_{N-1}   &h_0
\end{array}\right).
\end{equation}
As is well-known (see, e.g., Ref.~\cite{circulant}), the eigenvalues of the $H_{2d}$  matrix are given as
$E_{\tilde{s}}^{2d}=\sum_{m=0}^{N-1}h_m\rho_{\tilde{s}}^m$ with corresponding eigenvector $\frac{1}{\sqrt{N}}(1,\rho_{\tilde{s}},\rho_{\tilde{s}}^2,\ldots, \rho_{\tilde{s}}^{N-1})^T$, where $\rho_{\tilde{s}}$ is a complex number satisfying $\rho_{\tilde{s}}^N=1$.
For an integer ${\tilde{s}}$, here, we choose
\begin{equation}
\rho_{\tilde{s}}=\exp\left(i\frac{2\pi {\tilde{s}}}{N}\right),~~  -\frac{N}{2}\leq {\tilde{s}}< \frac{N}{2},
\end{equation}
 which is suitable both for  odd or even $N$. Hence, we have the energy eigenvalues of the $H_{2d}$  matrix:
\begin{equation}
E_{\tilde{s}}^{2d}= h_0 +\sum_{m=1}^{N-1} h_m e^{-2i\pi m{\tilde{s}}/{N}},
\label{eigen2d}
\end{equation}
and corresponding eigenfunction  to $E_{\tilde{s}}^{2d}$:
\begin{equation}
\phi_{\tilde{s}}^{2d}(x,y)=\frac{1}{\sqrt{N}}\sum_{j=0}^{N-1}e^{2i\pi j{\tilde{s}}/{N}}\phi_j(x,y),
\end{equation}
within the  approximation. As $\phi_j(x,y)$'s are real, we have 
\begin{equation}
h_m= <\phi_m|H_{2d}|\phi_0>=h_{N-m} ~~m=1,2,\ldots,N-1,
\label{N-m}
\end{equation}
implying $E_{\tilde{s}}^{2d}$  are real. Further, using Eqs.~(\ref{eigen2d}) and (\ref{N-m}), we  find
\begin{equation}
E_{\tilde{s}}^{2d}=E_{-{\tilde{s}}}^{2d}
\label{degenerate}
\end{equation}
which  shows that energy eigenvalues are degenerate if  $s$ is neither 0 nor $N/2$.
Using Eq.~(\ref{DRotation}), we also find that the eigenfunctions transform according to
\begin{equation}
\phi_{\tilde{s}}^{2d}(x_k',y_k')=e^{-2\pi i {k\tilde{s}}/{N} }  \phi_{\tilde{s}}^{2d}(x,y)
\label{Bloch}
\end{equation}
under the coordinate rotation of an integer $k$.

Considering the distance between the wells, it is natural to assume that $h_0\gg h_1=h_{N-1} \gg h_2=h_{N-2} \gg \cdots$. If we assume 
$h_2=h_{N-2}=h_3=h_{N-3}=\ldots=0$ as in Ref.~\cite{Sacchetti}, we arrive at
\begin{equation}
E_{\tilde{s}}^{2d}=h_0 +2h_1\cos\frac{2\pi {\tilde{s}}}{N}.
\label{tight}
\end{equation}
If $h_2, h_3, \ldots, h_{N-2}$ could be included within the approximation, then there will be additional corrections to $E_{\tilde{s}}^{2d}$ in Eq.~(\ref{tight}) which would also be sums of cosines.
In the large $N$ limit, the energy eigenvalues $E_{\tilde{s}}^{2d}$ form an energy band, and Eq.~(\ref{Bloch}) shows that  $\phi_{\tilde{s}}^{2d}(x,y)$ 
closely resemble the wave functions in Bloch (Floquet) theorem if the discrete rotations are corresponding to the translations by integral multiples of the period of the theorem. 
Further,   Eqs.~(\ref{degenerate},\ref{tight}) show that the lowest energy band of the 2-dimensional system is akin to those of the tight-binding case \cite{Harrison,AM}.








\end{document}